# Exploring the links between software development task type, team attitudes and task completion performance: Insights from the Jazz repository

Sherlock A. Licorish[✉], Stephen G. MacDonell

*Department of Information Science, University of Otago,
PO Box 56, Dunedin 9054, New Zealand*
sherlock.licorish@otago.ac.nz, stephen.macdonell@otago.ac.nz

**Abstract**

**Context**: In seeking to better understand the impact of various human factors involved in software development, and how teams' attitudes relate to their performance, increasing attention is being given to the study of team-related artefacts. In particular, researchers have conducted numerous studies on a range of team communication channels to explore links between developers' language use and the incidence of software bugs in the products they delivered. Comparatively limited attention has been paid, however, to the full range of software tasks that are commonly performed during the development and delivery of software systems, in spite of compelling evidence pointing to the need to understand teams' attitudes more widely. **Objective**: We were therefore motivated to study the relationships between task type and team attitudes, and how attitudes expressed in teams' communications might be related to their task completion performance when undertaking a range of activities. **Method**: Our investigation involved artefacts from 474 IBM Jazz practitioners assembled in 149 teams working on around 30,000 software development tasks over a three-year period. We applied linguistic analysis, standard statistical techniques and directed content analysis to address our research objective. **Results**: Our evidence revealed that teams expressed different attitudes when working on various forms of software tasks, and they were particularly emotional when working to remedy defects. That said, teams' expression of attitudes was not found to be a strong predictor of their task completion performance. **Conclusion**: Efforts aimed at reducing bug incidence may positively limit teams' emotional disposition when resolving bugs, thereby reducing the otherwise high demand for emotionally stable members. In addition, in environments where teams work closely together to develop software such as in Agile contexts, attitudes are likely to have a bearing on how they function as a group.

## 1. INTRODUCTION

Research addressing the management of software engineering[1] (SE) has recently cast a more focused lens on team-based issues, with in- creased attention on people and their work practices during software development and deployment [1–3]. In supporting the intent and execution of such work, repositories have played an increasingly important role, providing artefacts to enable various forms of team-focused explorations [4–7]. To date these works have considered a range of communication metrics and their linkages to software events and outcomes [6,8,9]. In particular, there has been a wealth of studies that have examined how teams' communication processes relate to the incidence of software bugs (see, for example, [10–13]). While these studies have indeed provided insights into the way software teams work, we argue that the body of evidence on SE management would benefit from explorations that address other software development activities, beyond just the study of bugs.

The need for and value of such work has long been recognized. In the 1980 s the work of McGrath and colleagues questioned the validity of outcomes derived from studies examining groups' and individuals' interactions and performance without considering the characteristics of the *full range of tasks* (defined in the next paragraph) that are commonly undertaken [14,15]. They claimed that the nature of the work in which a team is involved underpins the differences that are implicit in team tasks. Similarly, other early works have established that exploring teams' instrumental and expressive concerns and interpersonal needs provides a useful window for understanding their performance on various team-based endeavours [16].

In the general sense, software development essentially involves the writing of computer programs in response to client requests [17]. This process may of course involve several activities (or tasks), including those incorporated within feasibility studies, planning and documenting, coding, testing, bug fixing and deploying highly robust software [18]. After deployment the software may also be maintained for an extended period of time. Other tasks associated with software development include research, design, management, facilitation, and so on [19], and the writing of software itself may be supported by a range of tools, generators and the like. Tasks may be derived both implicitly, given a specific need arising from the detailed-level design or coding of a feature, and explicitly, in response to clients' requests [18]. Thus, there is a distinction between documented tasks (e.g., a client request) and those that are not documented but must be performed all the same to complete meaningful software development work (e.g., a programmer designing a prototype to evaluate the feasibility of developing a client feature). Tasks may also be aligned to multiple project objectives, and intertwined (e.g., coding a feature may also involve testing). Sometimes software tasks (other work) may also be performed during the detailed-level design and implementation of software projects that are not explicitly connected to the activities mentioned above (e.g., arranging software demonstration through site visits).

---

[1] Software engineering management involves the management of software development activities, including the design, coding, testing and maintenance of software products and services. Sometimes such activities may also extend to adapting and selling completed software products.

In this research we study those software engineering tasks that are largely explicit, with each defined as a unit of work associated with a particular project objective or feature (e.g., an enhancement request or new feature requirement). As with the range of software engineering tasks just mentioned, such work may take several weeks to complete, and require multiple practitioners (forming teams) to enable their implementation, while others may be delivered much more quickly and by a single practitioner.

While software engineering in general is accepted as being a com- plex undertaking due (in part) to the nature of the software product, which can be seen as conceptual and intangible in orientation [20], all software engineering tasks are not equal. For instance, coding a new feature or implementing feature enhancements is likely to necessitate relatively higher amounts of intellectual and cognitive processes [21], and such tasks will present a different level of difficulty, and will likely require different team configurations and idea generation processes [22], when compared to documentation, architectural design or support-related tasks. In contrast, these latter activities are likely to demand relatively higher levels of manipulative and cooperative requirements,[2] where social processes may feature more prominently, and particularly during consultation [16]. The need for high levels of cooperation when undertaking requirements gathering and software architecture design has indeed been observed by those examining the collaboration patterns of software developers [23]. Additionally, increased consensus [24] is likely to benefit those operating on documentation and software support and architectural design tasks. Similarly, teams resolving defects will likely require high degrees of familiarity and specific problem-solving knowledge (related to the previously developed feature), and so such tasks may demand less co-operation, tending towards smaller groups working with increased individual focus and employing intellectual processes [25].

Such differences are also likely to extend to the specific software development methods that are used. Agile methods, for instance, stress an iterative and fluid software development process, where software features are designed, constructed, and deployed in parallel, with a reduced emphasis on plans and processes [26,27]. These methods favour individuals and interactions, working software, customer collaboration and responding to change [28,29], all facets that would benefit from an understanding of teams attitudes. In contrast, more traditional software development processes emphasize a planned pro- cess, where typically the output of one phase of a project is used as input for the subsequent phase in progressing the project to completion using some variant of a waterfall process [20]. Given the growing use of Agile methods [29,30], and the emphasis placed on individuals and interactions in this context [31], it is necessary to understand how teams perform across various tasks in that context.

In fact, given that some of the above propositions have been established as plausible in the context of software engineering more generally, empirical analyses may generate tangible insights into the specific instrumental, expressive and interpersonal configurations that are most fitting for teams engaging in different forms of software engineering activities. These results could usefully inform software team composition strategies. Such insights could also provide in-depth contextual information that may enable researchers to further dissect how teams' attitudes relate to the incidence of software bugs. Our own prior study into how practitioners work while undertaking a range of software tasks indeed revealed that extrinsic factors (e.g., number of developers, number of messages practitioners communicated, the project phase in which the task is performed, and the task priority) accounted for less than 10% of the variance in task completion performance [32], suggesting that there are other influential factors at play in this context. Given our previous outcomes, we concluded that teams' attitudes and intrinsic issues may interact with these (and other) extrinsic variables in influencing software task completion performance. We therefore consider this issue in the work reported here.

We define task completion performance to be the duration taken to fulfil a software task obligation, measured by the number of day(s) it takes for a software task to be completed. However, in keeping with the generally fluid nature of software development, what is termed "complete" in software development is quite fuzzy. For instance, a task may be deemed completed today, and the software feature released, only for users to find bugs (or defects) months later and so the feature is in need of maintenance or repair. In some cases there would be no way to track the new work done to the task that was previously labelled as completed. This reality holds for all software projects. Thus, in this work we measure the completion of tasks based on their statuses as managed by software practitioners *up to the time of software release*: when completed, the status of a software task is set to resolved, closed or verified and the corresponding date added.

---

[2] Involving persuasion, negotiation and the need for consensus.

Our contributions are twofold. We first explore whether there is a link between the nature of the software tasks that teams are undertaking and the attitudes they express. We then examine how such attitudes are related to task completion performance. In addition to establishing the applicability of task differences, group work and organisational behaviour theories for studying teams engaged in software engineering tasks, we provide suggestions for those responsible for forming and managing software teams, and particularly those practitioners assembled to address specific subsets of software tasks. Furthermore, we assess the relevance of the expression of various attitudes for assessing team performance, in the process providing suggestions for managing software teams' instrumental, expressive and interpersonal facets (as defined in the section that follows).

In the next section we present our background and we survey related work. This leads to a statement of our specific research questions. We then provide details of our research setting and measures in Section 3, introducing our research techniques and procedures in this section. In Section 4 we present our results, and we discuss our findings in Section 5. In Section 6 we consider the threats to the validity of our study, and Section 7 outlines the implications of our results.

## 2. BACKGROUND AND RESEARCH QUESTIONS

In the general sense, attitude relates to an individual's way of thinking or the feelings they express about something [33]. We define team attitudes in this work as the aggregated sentiments expressed by the individuals that form a team [34,35]. Language use and the expression of sentiments have been used in past research to predict attitudes among small groups, and to explore how such attitudes when expressed in communication affect group dynamics [34]. We thus focus specifically on those sentiment types that are discernible in teams' communication [36,37]; being social, positive, negative, cognitive, work, and achievement. Social sentiments are communications that express various forms of affection, and positive sentiments are communications expressing joy and happiness [37]. Negative sentiments express discontent and unhappiness, while cognitive sentiments are communications expressing cognition and knowledge [35]. Work and achievement sentiments are communications focussed on task completion or accomplishment [38]. These sentiments are used collectively to study attitudes in this work. We provide further details of how the occurrence of these sentiments is measured in Section 3.2.

We provide our theoretical background in this section. The teams studied in this work used Agile-like practices in a distributed context. Thus, we first introduce Agile software development in Section 2.1, and we then review the literature on task differences and the ways in which these differences are perceived to influence individuals' and teams' attitudes in Section 2.2. We then consider the literature focused on how individuals' and teams' attitudes are related to their performance in Section 2.3, before formulating our research questions in Section 2.4.

### 2.1. Agile software development

The Agile Manifesto [31] promotes a software development philosophy that emphasizes a context in which the software development process is fluid and human-centred. Given the many voices that have lent their support to Agile methods these approaches have become widely adopted and studied [30]. Agile development is implemented by several methods, and according to Dybå et al. [39], of the many flavours of agile, Extreme Programming (XP) [26] and Scrum [40] are two of the most studied agile methods. Equally, these two approaches are also considered to be the most frequently adopted in the software development industry [41]. We thus briefly review these approaches in this section to contextualise our work.

The XP method is seen to implement the agile manifesto's [31] recommendations primarily through 12 practices, which are applied throughout a software development project. Concepts such as precise goals and a range of activities are central to this method (e.g., coding and testing). A major focus of the XP method is responding to changing customer requirements. This is facilitated through the development of software in small increments, and close customer collaboration. The on-site customer practice calls for the customer of the software development project to be always available. This practice is said to shorten feedback time, as developers can query the customer's opinion, and resolve issues rapidly, resulting in fewer costly readjustments [26].

Scrum's iterative and incremental framework stresses self-organisation around closely collaborating teams led by a Scrum Master. The Scrum approach is evidence-driven, where experiences and outcomes from earlier phases inform developers' plans and actions in subsequent phases. Scrum processes or ceremonies include sprint planning, daily Scrum (or stand-up), sprint reviews and retrospectives, whereas its artefacts include the product backlog, sprint backlog and burn down chart [40]. In the context of Scrum, customer feedback (as mentioned for XP) is largely implemented in the Sprint Review process, where the Product Owner sometimes serves as a customer representative and provides feedback [40].

### 2.2. Task differences and attitudes

A long-established body of work has considered task differences, and teams' attitudes when undertaking different forms of tasks. For instance, Carter, Haythorn, and Howell [21] classified team tasks into six types: clerical, discussion, intellectual construction, mechanical assembly, motor coordination and reasoning. The authors' position [21] was that each of these tasks has different performance processes, and so, their conduct demands different forms of team arrangement. McGrath and Altman's [15] subsequent review of small group research provided a similar classification to that of Carter, Haythorn, and Howell [21], and emphasized the necessity of particular team orientations for the execution of specific tasks. Subsequent work by Shaw [16] extracted six dimensions of group tasks when examining previous work: (1) intellective versus manipulative requirements, (2) task difficulty, (3) intrinsic interest, (4) population familiarity, (5) solution multiplicity versus specificity, and (6) cooperation requirement. While the first of Shaw's dimensions [16] refers to the property of a particular task, others consider the relation between the task and those performing the task (i.e., dimensions 2, 3 and 4 respectively), how the task is evaluated (dimension 5) and how individuals must act while working together to achieve the task outcome (dimension 6). Beyond Shaw's broad classification, Hackman [25] and Hackman, Morris, and Leonard [42] created a task model focused on the intellectual tasks that led to written products. Their research revealed three forms of task: production tasks (tasks requiring idea generation or creativity), discussion tasks (requiring dialog), and problem solving tasks (involving the execution of a plan). These studies and their models all lend credence to the general necessity to understand task differences when evaluating teams' performance.

Of more specific relevance to this work are the variances in attitudes that arise when teams undertake various forms of task. Such attitudes are likely to be influenced by the demeanours of the team's individual members. The members' conduct in turn is linked to their individual properties (traits, characteristics, beliefs and habits) [14]. Thus, enquiries considering team attitudes are often encouraged to also take *individual* members' dispositions, as well as group structure, into account. More importantly, given that group work is performed in specific contexts, and is focused on particular tasks, this additional variable (i.e., the task itself) is likely to be related to the attitudes that are expressed by teams [14]. Furthermore, as noted above, the task itself may involve particular complexities, and so these details might also need to be taken into account when studying teams' attitudes. It has been noted that group member properties as granular as age, gender, disposition, belief, moods, state of mind and motives, along with aspects of the group's physical environment (e.g., noise, heating, lighting) may affect both the attitudes expressed by teams and their performance [14]. In fact, this list represents only a modest subset of those variables that might be of influence, which may in fact be infinite in number. However, there is little prior theoretical support for consideration of the majority of these variables.

Notwithstanding the vast array of potentially important variables, there is theoretical support for the position that specific task demands impact team interactions and the way teams perform in particular situations [43]. Beyond the expression of sentiments above, attitudes here may also be granularly defined in terms of rates of interaction, the distribution of participation and members' involvement, the flow of information and the flow of interpersonal effect. Such attitudes would be especially noteworthy in Agile software development environments, where individuals and interactions are central to teams' ways of working. Hence, studies aimed at examining the attitudes of Agile teams would likely provide useful insights for this community.

**2.3. Attitudes and task completion performance**
The attitudes expressed by individuals working together are believed to impact their team's ability to successfully complete assigned tasks. While all activities conducted by teams result in instrumental (work and task focus) and expressive (social-emotional and interpersonal focus) attitudes, increased task difficulty is also said to exert more strain on teams, throwing out the social-emotional balance that is necessary for positive task performance during group work. Accordingly, sometimes teams must neglect instrumental concerns to commit exclusively to expressive concerns in order to reduce tension and support team morale [14]. This may lead to undue delays (or reduced effort on tasks), but is often necessary for the long-term functioning of the team. Bale [44] notes that positive team attitudes must exceed negative team attitudes if teams are to complete tasks success- fully. Negative team attitudes lower team motivation, and so there is often a need for higher levels of positive team attitudes if teams are to maintain the average level of satisfaction necessary to complete as- signed tasks. In other words, if negative and positive team attitudes are equivalent during team problem-solving the team's level of motivation is not likely to lead to successful task completion [14]. Bale [44] indeed found that groups with higher levels of positive-negative attitude ratios have higher levels of satisfaction. Dittes and Kelly [45] also found that rejected members reduced their communication, while Pepinsky et al. [46] observed that the opposite conduct was exhibited by members who were positively supported.

Others have provided slightly different models to explain this phenomenon; however, these models can all be interpreted under Bale's two-process schema. For instance, Thelen [47] used group theory to provide a model encompassing a similar distinction of work and emotion to that of Bale. However, he classified work under four sub-dimensions and emotion under three. Additionally, Thelen's model outlined that both work and emotion are expressed in one communication act. Hare's model [48] considers four group functions that may also be classified under Bale's two-process model relating to task execution and interpersonal relations. Others have considered this issue from a group development perspective. For example, the work of Schutz [49] uncovered three sequential interpersonal needs: inclusion, control and affection. Schutz asserted that cycles of interpersonal action reoccur throughout the early phases of group work, but then reverse (i.e., team members break ties of affection, release control and then stop interacting to release group identity) towards task completion. Tuckman's seminal work [50] on various forms of groups uncovered that they traverse four major stages, with each stage of development comprising group structure and task aspects. Stage one (forming) normally involves testing the group structure and the task and attitudinal aspects of the team, while intra-group conflicts around tasks emerge in stage two (storming). Groups generally tend to be more cohesive and open in stage three (norming), and the team becomes more functional and insightful in stage four (performing). Others have uncovered similar patterns to those promoted by Tuckman during the study of group development, for example, Mills [51] and Slater [52]. These works have all confirmed that there is a relationship between a team's expression of attitudes and their group processes and performance, which warrants the consideration of this issue in an Agile software engineering setting.

## 2.4. Research questions

While previous work in software engineering has provided a wealth of insights into how software teams work [5,53], particularly when engaged with bug issues [12], less attention has been given to the other forms of software engineering tasks. On the backdrop of evidence that multiple properties of team tasks affect team performance [14,15,25], it seems reasonable to explore if teams express different attitudes when undertaking different forms of software task. If this insight is established, it would provide a platform for future work to explore if teams are likely to benefit from particular pre-set configurations and team arrangements. In fact, exploring the way developers work and their motivations across the range of software tasks that are commonly performed, and particularly those initial actions that lead to the development of software features in the first place, and then subsequent bugs, could provide added value for the software engineering community.

Along these lines, the wider body of evidence around software team attitudes and affective factors in software engineering has provided a range of explanations for the way teams conduct their activities [54], on the basis that developers express emotion that are captured in their artefacts [55]. In particular, previous works have paid particular attention to developers' communications [56,57]. For instance, Pletea et al. [58] gauged atmosphere surrounding security-related discussions on GitHub as mined from discussions around commits and pull requests, finding a higher incidence of negative emotions in security-related discussions than in other discussions. Developers' language use was also shown to be related to the priorities assigned to issues [59]. Furthermore, attention was also specifically given to negative affect given perceptions that negative team processes may lead to undesirable outcomes [60]. For instance, anger has been considered by Gachechiladze et al. [61], who focussed on hostility and resentment towards self, others and objects. Their evidence shows that developers' anger was expressed predominantly towards tools and programming languages. It is assumed that such understandings could lead to improved team management and the provision of support to overcome challenges related to team conflicts and behavioural differences. In fact, efforts have also been made towards developing tools to aid with visualisation of team moods and sentiments in order to monitor team climate [62,63].

The previous body of work focussed on human factors in software engineering has provided a range of circumstantial evidence around how software teams behave. However, less attention has been given to the range of software engineering tasks that are commonly performed. We thus propose to answer our first research question in examining artefacts from Agile teams:

> RQ1. What attitudes do team members express when undertaking different forms of software task?

Previous studies have observed that many factors influence software teams' delivery performance (e.g., the number of developers, feature size and response time) [6,64]. However, we did not find significant relationships between these variables and teams' task completion performance in an earlier study [32]. As a result, and in line with our review of the related literature reported above [14,45,48], we anticipate that teams' instrumental and expressive concerns and interpersonal needs may interact with extrinsic task variables in influencing software tasks' completion performance [44]. Although previous work has examined practitioners' and teams' attitudes [65,66], and has related such attitudes to their involvement in development activities [67–69] and project governance [65,70] (e.g., how many messages they communicate), limited attention has been given to studying the way software teams' attitudes covary with their tasks' completion performance, considering their full range of tasks.

That said, there is adequate evidence to suggest that such understandings would be valuable for the software engineering community, and particularly for instances where negative attitudes may be evident. For instance, Graziotin et al. [71] explored the consequences of unhappiness among software developers, finding 49 consequences of un- happiness while undertaking software development, with impacts on productivity and performance being particularly pronounced. Others have linked unhappiness to negative effects both for developers personally (e.g., their well-being) and on development outcomes (e.g., product quality), with being stuck in problem solving and time pressure being the two most frequent causes of unhappiness [72]. In contrast, developers expressing higher levels of positive emotion and politeness were shown to take less time to address issues, which in turn resulted in happiness [73].

With Agile teams working closely to deliver software tasks in highly collaborative environments, the balancing of attitudes would seem to be necessary. Thus, evidence for the way attitudes covary with task completion performance would be enlightening, in terms of providing support for Agile team composition strategies. In fact, as noted in Section 2.2 above, a range of other factors related to group structure [74], situation-related [6], and interaction-related aspects [75,76] may also affect task completion performance, and so, such factors should be controlled when examining how teams' attitudes covary with software task completion performance. We thus outline our second research question to direct this enquiry:

> RQ2. How do the attitudes expressed by the team covary with software task completion performance?

## 3. RESEARCH SETTING

To answer our research questions we used linguistic and directed content analysis techniques to examine artefacts produced during three years of development of Jazz 1.0.1 (based on the IBM[R] Rational[R] Team Concert (RTC)[3]). Jazz is a fully functional environment for developing software and for managing the entire software development process [77]. The software includes features for work planning and traceability, software builds, code analysis, bug tracking and version control in one system [78]. Changes to source code in the Jazz environment are permitted only as a consequence of work items (WIs) being created beforehand, such as a defect, a [support] task or an enhancement request. Defects represent work related to bug fixing, whereas design documents, documentation or support for the RTC online community are labelled as tasks (although we refer to them here as 'support tasks' in order to differentiate with our general use of 'task' in the paper). Enhancements relate to the provision of new functionality or the extension of system features. These are all explicit software development tasks (refer to Section 1 for details). Team members' communication and interaction around WIs are captured by Jazz's comment or message functionality. During development at IBM, project communication, the content explored in this study, was enforced and captured through the use of Jazz itself [6].

---

[3] IBM, the IBM logo, ibm.com, and Rational are trademarks or registered trademarks of International Business Machines Corporation in the United States, other countries, or both.

The Jazz repository thus comprised a large amount of process data collected from distributed software development and management activities across the USA, Canada and Europe. Teams were resident in each location, and worked jointly as a wider group to develop Jazz and RTC related products. It was entirely possible, of course, that communication and interaction also occurred through other means outside Jazz, in cases where developers felt constrained by the system; however, we do not have access to these communications. Thus, we primarily study the communication as captured in Jazz (refer to Section 6 where we address this limitation).

In Jazz each team is made up of multiple individual roles, with a project leader responsible for the management and coordination of the activities undertaken by the team [79]. All Jazz teams use the Eclipse-way methodology for guiding the software development process [77]. This methodology is similar to the Open Unified Process (OpenUP[4]), and outlines iteration cycles that are six to eight weeks in duration, comprising planning, development and stabilizing phases, and generally conforming to the agile principles. Development (in iterations) is driven by work items, which are the most granular unit of work. Teams collaborate around these work items by sharing understandings and balancing competing priorities, where an early focus on the software architecture is typical before the project evolves with continuous feedback. Builds are executed after project iterations. All information related to the software process is stored in a server repository, which is accessible through a web-based or Eclipse-based (RTC) client interface [80]. This consolidated data storage, coupled with high levels of project control, mean that the data in Jazz is much more complete and representative of the software process than that in many OSS repositories. Thus, replicating the work that is performed here for OSS projects would not necessarily lead to comparable outcomes, given the differences in quality control for Jazz and other OSS projects. For example, whereas in Jazz each work item is labelled, addressed a particular unit of work, and specific identifiable team members are assigned to solve work items, features are often not as clearly classified in OSS projects and it is also quite challenging to identify the members that developed specific features.

We extracted the Jazz data and applied linguistic analysis, standard statistical techniques and directed content analysis to answer the two research questions introduced in Section 2.4. We provide details of our data extraction process and metrics definitions in the following two subsections.

### 3.1. Data extraction

We briefly report here the aspects of data mining that supported the activities involved in this project in terms of extracting, preparing and exploring the data under observation [81]. Data cleaning, integration and transformation techniques were utilized to maximize the representativeness of the data under consideration and to help with the assurance of data quality, while exploratory data analysis (EDA) techniques were employed to investigate data properties and to facilitate anomaly detection [82]. Through these latter activities we were able to identify all records with inconsistent formats and data types, for example: an integer column with an empty cell. We wrote scripts to search for these inconsistent records and tagged those for deletion. This exercise allowed us to identify and delete 122 records that were of in- consistent format. We also wrote scripts that removed all HTML tags and foreign characters (as these would have confounded our analysis).

We leveraged the IBM Rational Jazz Client API to extract team information and development and communication artefacts from the Jazz repository. In total we extracted 30,646 resolved WIs (labelled as one of the three types described above) developed by a total of 474 contributors working on these tasks between June 2005 and June 2008. These contributors belonged to five different roles: Team leads (or component leads) are responsible for planning and executing the architectural integration of components; Admins are responsible for the configuration and integration of artefacts; Project managers (PMC) are responsible for project governance; those occupying the Programmer (contributor) role contribute code to features; and finally, those who occupied more than one of these roles were labelled Multiple. The features were divided among 149 functional teams, and 117,101 messages (comments) were exchanged in relation to these WIs. Some teams in our particular snapshot worked on as little as one WI, while the maximum number of WIs assigned to one team was 4851. These WIs were developed across 30 iterations, where iteration cycles were six to eight weeks in duration.

As noted above, the Jazz project teams were employed across locations in North America and Europe; however, we did not consider the specific team location as a relevant unit of analysis in this work. We are aware that cultural differences and distance (geographical and temporal) may affect software development teams' performance [83], and these conditions may also have an impact on team members' attitudes - which in turn may lead to performance issues [4]. However, previous research examining the effects of cultural differences in global software teams has found few cultural gaps and differences in attitudes among software teams operating in Western cultures (the setting for the teams studied in this work). The largest negative effects between global teams were observed between Asian and Western cultures [83]. Accordingly, in this study we focus on a number of other control factors as derived from the literature around task differences, organizational behaviour and group work in general [75]. These factors, along with the other study measures used are described next.

---

[4] https://eclipse.org/epf/general/OpenUP.pdf.

Table 1. Coding categories for exploring teams' interactions.

| Category | Characteristics and example |
| --- | --- |
| Work and Achievement Comments | Share information – "Just for your information, we successfully integrated change 305 last evening." |
| | Provide guidance and suggestions to others – "Let's document the procedures that were involved in solving this problem 305, it may be quite useful." |
| Negative Comments | Judgmental – "I disagree that refactoring may be considered the ultimate test of code quality." |
| Cognitive Comments | Elaborate, exchange, and express ideas or thoughts – "What is most intriguing in re-integrating this feature is how refactoring reveals issues even when no functional changes are made." |
| Social and Positive Comments | Thankful or offering commendation – "Thanks for your suggestions, your advice actually worked." |
| | Communication not related to solving the task under consideration – "How was your weekend?" |

### 3.2. Description of measures

We applied linguistic and statistical analysis methods to a range of metrics computed from the extracted Jazz data to answer our questions. The software task was our unit of analysis in this work, and practitioners collaborating around a particular software task comprised a team. Tasks were already categorized in the repository, and we explain how these were analysed below. Team members' attitudes and task completion performance formed our dependent variables, and a number of control factors were also included in our analysis so that we might more fully understand task completion performance. We now examine each of these variables and how they were operationalized in turn.

### 3.2.1. Measuring task type

As noted above, each software task is categorized as a defect, an enhancement or a support task in the Jazz repository. Although these tasks are broadly classified, quite specific work is captured under each category. During Jazz development, tasks that were categorized as defects were focused on bug fixes, while WIs that were labelled as support tasks captured work on design documents, documentation or support for the RTC online community [79]. Tasks that were classified as enhancements related to the provision of new software functionality or the extension of established system features. It was therefore straightforward to use the task classification scheme from the repository to group and examine the Jazz software teams' undertakings.

### 3.2.2. Measuring attitudes

We used linguistic and directed content analyses to measure and study teams' attitudes. We introduced the operationalisation of attitudes in Section 2, but here extend these discussions by considering the use of our linguistic and directed content analysis techniques in the following two subsections.

*(i) Linguistic analysis:* Language use has been studied extensively across a range of social contexts [36,37,84-86], and so was considered suitable for studying attitudes in this research. These works have all provided evidence in support of the contention that there are unique variations in individuals' (and teams') linguistic styles from situation to situation, and that linguistic analysis of textual communication can reveal the attitudes that are expressed by those who are communicating. In following the lead of previous work [87,88], we employed the Linguistic Inquiry and Word Count (LIWC) soft- ware tool in our analysis of practitioners' and teams' attitudes. The LIWC tool was created after four decades of research using data collected across the USA, Canada and New Zealand [37]. This tool captures over 86% of the words used during conversations. Written text is submitted as input in a file that is then processed and summarized based on the LIWC tool's dictionary. Each word in the file is searched for in the dictionary, and specific type counts are incremented based on the associated word category (if found), after which a percentage value is calculated by aggregating the number of words in each linguistic category over all words in the messages. For example, if there were 10 instances of words belonging to the social dimension (as defined below) in a message with a length of 200 words then the percentage value for the social dimension would be (10/200 = )5.0%. The dimensions in the LIWC output summary are said to capture the attitudes of those communicating as reflected in the words they use [37,88]. In assessing team attitudes we selected six classes of attitudes that can be readily detected in language use to assess interpersonal processes: *social, positive, negative, cognitive, work,* and *achievement* attitudes. The *social, positive,* and *negative* dimensions were used to study teams' interpersonal focus and positive-negative attitudes ratios (considered in Section 2.3). Teams' instrumental concerns were examined using the *cognitive, work and achievement* linguistic dimensions. To briefly illustrate, a social attitude is indicated through the use of words such as "give", "buddy" and "love", while words including "think", "consider" and "determine" reflect *a cognitive* attitude [37].

*(ii) Directed content analysis (CA):* We triangulate our LIWC findings through an in depth examination of 1261 messages that were contributed in relation to 250 randomly selected software tasks using a directed content analysis approach. We employed a hybrid classification scheme adapted from prior works that had examined the details of teams' interactions. The classification schemes of Henri [89] and Zhu [90] are particularly applicable to the work undertaken in this research because of their treatment of interaction – the study of which is said to reveal if teams express different attitudes when undertaking different forms of software task.

Use of a directed CA approach is appropriate when there is scope to extend or complement existing theories around a phenomenon [91], and so suited our further explorations of

Table 2. Summary of measures.

| Variable or grouping | Description or sub-categories | Category of variable(s) | Research question(s) |
| --- | --- | --- | --- |
| Software task (type) | support, enhancement and defect | Independent | RQ1 |
| Attitudes* | social, positive, negative, cognitive, work, achievement | Dependent, Independent | RQ1, RQ2 |
| Task completion performance | number of days (time taken) | Dependent | RQ2 |
| Group structure | number of developers, number of roles, individual roles | Control | RQ2 |
| Situation-related factors | task priority, iteration | Control | RQ2 |
| Interaction-related factors | Number of comments, message length (number of words) | Control | RQ2 |

Jazz teams' expression of attitudes when undertaking different forms of task. The directed content analyst approaches the data analysis process using existing theories to identify key concepts and definitions as coding categories. In our case, we used theories examining knowledge-sharing expressions in textual interaction [89,90]. Relevant categories with appropriate examples are included in Table 1. Henri [89] and Zhu [90] used Bretz's [92] three-stage theory of interactivity and the group interaction theory of Hatano and Inagaki [93] and Graesser and Person [94] respectively to study teams' interactions. Henri's [89] coding instrument was created to observe five dimensions of interactivity: participative, social, interactive, cognitive and meta-cognitive communication, while Zhu's [90] social interaction protocol looked to classify vertical or horizontal interaction. Vertical interaction is characterized by communication where group members seek answers or solutions to problems from (more) capable members, while horizontal interaction involves the strong assertion of ideas, answers, information, discussions, judgement exchanges, reflections and scaffolding. Given the focus of this research we were particularly interested in comments that were related to work and achievement, as well as those that were negative or judgmental, cognitive, and social and positive in nature (refer to Table 1).

In evaluating the categorization in Table 1, the authors and two other experienced coders first classified a random sample of 5% of the 1261 comments and found that members in Jazz communicated multiple ideas in their messages, and so some utterances demonstrated more than one form of interaction. We thus segmented the communications using the sentence (or utterance) as the unit of analysis, after which the first author and the two experienced coders coded the 1261 messages that were communicated based on the protocol in Table 1. Multiple codes were assigned to utterances that demonstrated more than one form of interaction, and all coding differences were discussed and resolved by consensus. In addition, we were only focused on codes that matched the categories in Table 1, and thus, while others were aggregated (refer to Section 4), their specific details are not reported in our findings. We achieved 81% inter-rater agreement between the three coders as measured using Holsti's coefficient of reliability measurement (C.R.) [95]. This represents excellent agreement between coders and suggests that a consistent and reliable approach was taken.

### 3.2.3. Measuring task completion performance

Various approaches have been used over many years to measure team- and individual-level performance while undertaking software development tasks. Productivity-related measures such as lines of code per unit of effort [96] and the number of task changes completed [97] are among those that have been used previously to measure performance. Cataldo and Herbsleb [97] argued that measures based on lines of code may not be reliable in instances where there is variability in developers' coding styles (e.g., some developers are more verbose than others). Additionally, although task changes may be useful for studying performance at the individual practitioner level [7], this metric is not suitable for studying *task completion performance* by teams.

Therefore, given the use of the task as the unit of analysis in this work, performance was most appropriately measured at the task level [98], for tasks that had been completed. We therefore computed the task completion performance of each task by calculating the number of day(s) it took for the task to be completed. A task was assessed as completed if the status was set to resolved, closed or verified and the corresponding date added. Such an approach has also been used extensively by others to measure performance [79,99]. We also considered a number of control variables, including those related to group structure, situation-related factors and interaction-related factors, to examine the way teams' expressions of attitudes covary with task completion performance, as introduced next.

### 3.2.4. Control factors/variables

Of the intervening variables that are measurable, properties of group structure, situation-related and interaction-related factors are considered to be potentially important when assessing the outcomes of collective action [15,21,25,42,79,100]. Thus, we organize our *control variables* along these three dimensions when studying software task completion performance.

*(i) Group structure:* Jazz developers were assigned to one of four distinct roles noted above (while those occupying multiple roles were assigned to the fifth role *Multiple* in this study). Given that such roles were assigned by upper level management and that specific intrinsic responsibilities may be assigned to these roles [79], coupled with previous evidence that has established that members' status may impact task outcomes [74], we considered the distribution of roles as potentially related to task completion performance. We aggregated the number of developers, the number of unique roles, and the number of individual roles, respectively, to measure *group structure*.

*(ii) Situation-related factors:* Tasks with higher priority should be completed sooner than those that are considered as less critical [6].

Similarly, tasks developed in certain phases may be completed with greater urgency than those that are less urgently needed (e.g., those features that are worked on closer to a delivery date are likely to be done with greater urgency than those developed at the start of the iteration). We therefore considered the priority of the task and the iteration in which the task was created as *situation-related* control factors during our assessment of task completion performance.

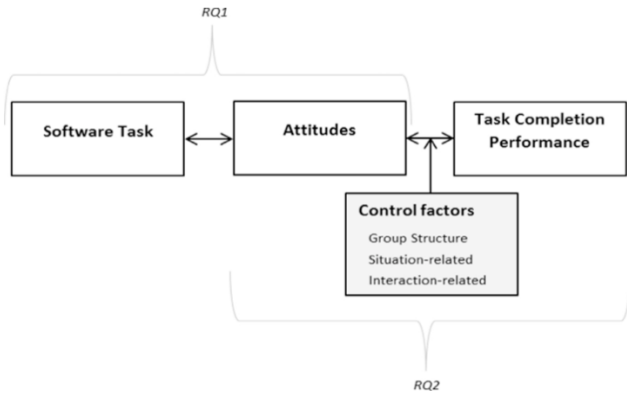

Fig. 1. Study model depicting relationships between software tasks, teams' expressions of attitudes, classes of control factors, and task completion performance.

*(iii) Interaction-related factors:* In line with previous work [64], we categorized a number of communication-related control factors that are associated with team engagement and participation, and particularly, those related to communication structures which may impact task completion performance. For instance, information diversity has been shown to help with task innovativeness [75,76], and diversity also enhances competitiveness [101]. However, the need to manage a very large volume of information also results in information overload and task delays [102,103]. Under both circumstances task completion performance may be affected. We therefore considered the number of comments and the volume of words communicated in messages around software tasks as potentially impacting on task completion performance. These metrics were accommodated under *interaction-related* control factors.

Our variables are summarised in Table 2, and Fig. 1 provides a pictorial representation of our conceptual model. We first consider whether teams expressed different attitudes when undertaking different forms of software task in answering RQ1. Taking a range of control variables into consideration, we then address RQ2, in examining whether the attitudes expressed by the team covary with software task completion performance (described above). We present our results in the next section.

## 4. RESULTS

Of the more than 30,000 software tasks that we extracted from the Jazz repository the largest single group consisted of 23,331 tasks (76.1%) classified as defects. A further 12.2% (3748 of the 30,646 tasks) were classified as support tasks, and the remaining 11.6% (3567 tasks) were classified as enhancements (providing new functionality or the extension of system features). While there was an overlap in team members undertaking each form of task, overall, defects attracted the largest cohort of members (411 practitioners or 86.7% of the 474 total members), enhancements attracted the second highest number of team members (226 practitioners or 47.7%), and support tasks involved the fewest members (212 practitioners or 44.7%). We provide summary descriptive statistics for the Jazz dataset in Table 3. Here it is evident that on average there were around two team members working on each task, regardless of the type (mean, enhance features = 2.0, defect = 2.0 and support tasks=1.8). Of the 117,101 messages that were exchanged around the 30,646 tasks, 88,874 messages (or 75.9%) were exchanged by teams working on defects, 14,512 messages (12.4%) were exchanged by teams working on enhancement features, and 13,715 messages (11.7%) were exchanged by teams working on support tasks. Table 3 shows that teams exchanged most messages when they were working on enhancement features (number of messages exchanged, mean=4.1, median=3.0, std dev=4.8), with slightly fewer messages communicated around defects (mean=3.8, median=2.0, std dev = 4.5), while the fewest messages were exchanged around support tasks (mean = 3.7, median = 2.0, std dev = 4.5). We also observe in Table 3 that the distributions for the three forms of task were skewed, with both skewness and kurtosis values positively oriented. We analysed this comprehensive snapshot of rich data using linguistic, statistical and content analysis techniques to answer the two questions out-lined in Section 2; our results for each are provided in the following two subsections.

*RQ1. What attitudes do team members express when undertaking different forms of software task?*

We first analysed the messages that were shared among teams working on the three forms of tasks according to the linguistic dimensions (*social, positive, negative, cognitive, work,* and *achievement*) outlined in Section 3.2, then computed descriptive statistics to explore any variations across these dimensions. A summary of these statistics and mean ranks is provided in Tables 4 and 5; this shows that there were indeed variations in linguistic usage for the different teams undertaking the three forms of software tasks. Most notable in Tables 4 and 5 are the differences for the *negative, work* and *achievement* dimensions. We employed formal statistical techniques to assess the differences in teams' linguistic usage across the three forms of tasks. Given the large sample size noted above, we first used a series of Kolmogorov-Smirnov tests to check the normality of teams' linguistic usage. The results of these

Table 3. Descriptive statistics for the extracted Jazz dataset.

|           | Task    | Mean | Median | Std Dev | Min | Max | SE (Mean) | SK   | KS   | SE (SK) | SE (KS) |
|-----------|---------|------|--------|---------|-----|-----|-----------|------|------|---------|---------|
| Members   | Support | 1.8  | 1      | 1.2     | 1   | 13  | 0.02      | 2.5  | 10.5 | 0.04    | 0.08    |
|           | Enhance | 2    | 2      | 1.2     | 1   | 10  | 0.02      | 1.9  | 5.6  | 0.04    | 0.08    |
|           | Defect  | 2    | 2      | 1.2     | 1   | 19  | 0.01      | 2.0  | 8.8  | 0.02    | 0.03    |
| Exchanges | Support | 3.7  | 2.2    | 4.5     | 1   | 74  | 0.03      | 4.4  | 33.5 | 0.04    | 0.08    |
|           | Enhance | 4.1  | 3.0    | 4.8     | 1   | 57  | 0.08      | 3.9  | 23.2 | 0.04    | 0.08    |
|           | Defect  | 3.8  | 2.0    | 4.5     | 1   | 266 | 0.07      | 11.5 | 50.0 | 0.02    | 0.03    |

*Note*: Std Dev = Standard deviation, SE = Standard error, SK = Skewness, KS = Kurtosis.

Table 4. Descriptive statistics for attitudes for different forms of software task.

| Linguistic dimension | Mean | | | Median | | | Std Dev | | |
|---|---|---|---|---|---|---|---|---|---|
| | Sup. | Enh. | Def. | Sup. | Enh. | Def. | Sup. | Enh. | Def. |
| Social | 3.5 | 3.8 | 3.1 | 2.4 | 3.1 | 2.3 | 4.4 | 3.9 | 3.7 |
| Positive | 4.0 | 4.2 | 4.5 | 1.5 | 2.1 | 2.1 | 7.4 | 7.1 | 7.1 |
| Negative | 0.7 | 0.7 | 1.2 | 0.0 | 0.0 | 0.0 | 2.0 | 1.7 | 2.7 |
| Cognitive | 12.0 | 12.4 | 11.6 | 12.0 | 12.8 | 11.8 | 9.2 | 8.7 | 8.1 |
| Work-focused | 5.7 | 4.8 | 4.0 | 3.7 | 3.3 | 2.6 | 6.7 | 5.3 | 4.9 |
| Achievement-focused | 5.3 | 4.5 | 3.6 | 3.2 | 3.0 | 2.3 | 6.5 | 5.2 | 4.7 |

Note: Sup. = Support, Enh. = Enhance, Def. = Defect.

Table 5. Mean ranks and Chi-Square values for attitudes for different forms of software task.

| Linguistic dimension | Mean rank | | | Chi-Square |
|---|---|---|---|---|
| | Support | Enhance | Defect | |
| Social | 15,503.0 | 16,755.8 | 15,075.7 | 118.4 |
| Positive | 14,200.8 | 15,398.5 | 15,492.4 | 71.8 |
| Negative | 13,462.2 | 14,122.2 | 15,806.2 | 368.9 |
| Cognitive | 15,386.1 | 16,067.8 | 15,199.7 | 30.2 |
| Work-focused | 16,823.1 | 16,390.7 | 14,919.4 | 212.5 |
| Achievement-focused | 16,773.8 | 16,518.8 | 14,907.8 | 222.6 |

tests confirmed that the data distributions for all six linguistic categories (*social, positive, negative, cognitive, work,* and *achievement*), for each of the three task types (support tasks, enhancements, defects), significantly deviated from a normal distribution (p < .05). The standardized coefficients for skewness and kurtosis (i.e., the skewness and kurtosis values divided by their respective standard errors) were also outside the boundaries of normally distributed data (i.e., −3 to + 3) [104]. Thus, a series of six nonparametric Kruskal-Wallis tests was used to test for differences in teams' linguistic usage (for *social, positive, negative, cognitive, work,* and *achievement*), when working to complete the three forms of tasks (support tasks, enhancements, defects). These tests revealed statistically significant differences in the way teams expressed themselves while undertaking support, enhancement and defect tasks, for all six linguistic dimensions (p < .01 for all observations). Measures returned for the Kruskal-Wallis tests for mean ranks and Chi-Square values are provided in Table 5, showing that teams were particularly *social* and *cognitive* when working to complete enhancements (higher scores correspond with higher means in Table 4). Table 5 also shows that teams were most negative when resolving defects, and they expressed more *work* and *achievement* focus while working to complete both support and enhancement tasks.

Given the findings reported in Tables 4 and 5, a series of Mann-Whitney pair-wise follow-up tests at the Bonferroni adjusted level of 0.016 (i.e., 0.05 divided by 3 analyses – representing the three forms of tasks) were performed to test for pair-wise differences in usage of the six linguistic dimensions (*social, positive, negative, cognitive, work,* and *achievement*) across the three forms of tasks (support tasks, enhancements, defects). These results indicated that teams were significantly more social when completing support and enhancement tasks than when working on defects (p < .016 for each comparison), and a statistically significant finding was also revealed when a pair-wise comparison was conducted for support and enhancement tasks (p < .016). Pair-wise comparisons also revealed that teams were significantly more *positive* when addressing enhancement and defect tasks than when completing support tasks (p < .016 for each comparison). Teams expressed significantly more *negative* language when fixing defects than when they were occupied on the other forms of tasks (p < .016 for each comparison), and team members expressed more negative language when completing enhancements than when they were undertaking support tasks (p < .016). Teams were also significantly more *cognitive* when working towards enhancements than when conducting support and defect tasks (p < .016 for each comparison), whereas *work* and *achievement* focus were significantly higher among teams working to complete support and enhancement tasks than defects (p < .016 for each comparison).

We normalized our data and performed another round of analysis to triangulate these results. We first selected linguistic measures for individual team members who worked on and communicated in relation to all three forms of task in their teams, to examine differences in their attitudes. Of the 474 total members, 152 met this criteria (i.e., they worked on all three forms of task at some time and submitted messages about their tasks). We replicated the Kruskal-Wallis tests noted above for the six linguistic dimensions *(social, positive, negative, cognitive, work,* and *achievement)* across the three forms of tasks (support tasks, enhancements, defects). Results from the six tests revealed that, with the exception of social expression, these members expressed significantly different attitudes when they were working to complete the three different forms of task (p < .01 for the five comparisons – the exception being for social linguistic utterances). Mean ranks for the usage of the six linguistic dimensions are provided in Fig. 2, and follow up Mann-Whitney pair-wise tests for linguistic usage for the five dimensions where there were statistically significant differences (positive, negative, cognitive, work, and achievement) across the three forms of tasks also confirmed the pattern of results just noted.

Our directed content analysis results are next examined. Of the 250 work items selected, 150 were defects, 50 were enhancements and 50 were support tasks. As noted above, teams communicated 1261 messages around these three forms of tasks: 738 messages relating to defects, 294 messages for the support tasks, and 229 messages around enhancements. From these 1261 messages, 3630 codes were recorded to the categories in Table 1 (support = 800, enhancement = 745 and defect = 2085). Table 6 provides summary counts of these codes. Given the differences in the number of messages that were coded for the three respective tasks we normalize the outcomes in Table 6 to compare our outcomes with those from the LIWC tool above. Fig. 3 depicts these normalized scores, which shows a similar pattern to the results in Table 4 for social and positive (support = 4.9%, enhancement = 4.8% and defect = 5.3%) and cognitive utterances (support = 27.5%, enhancement = 21.9% and defect = 18.9%). Negative utterances remain prominent for defects (8.7%), but we see in Fig. 3 that there was also some degree of negative utterances expressed around enhancements (9.4%). This form of utterance remained consistent with the linguistic outcomes in Table 4 for support tasks (8.1%). However, in Fig. 3 the outcomes for work and achievement utterances are somewhat divergent to the linguistic outcomes in Table 4 (support = 59.5%, enhancement = 63.9% and defect = 67.1%). Notwithstanding that we only coded 1261 of the total 117,101 messages (comments), our directed content analysis outcomes largely converge with those obtained from our analysis of the LIWC tool output.

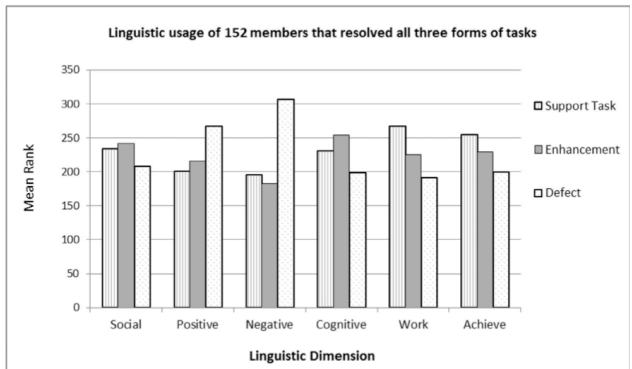

Fig. 2. Mean ranks for teams' linguistic usage.

Table 6. Interaction categories and number of occurrences for support, enhancement and defect tasks.

| Category | Support | Enhance | Defect |
|---|---|---|---|
| Work and Achievement Comments | 476 | 476 | 1399 |
| Negative Comments | 65 | 70 | 182 |
| Cognitive Comments | 220 | 163 | 394 |
| Social and Positive Comments | 39 | 36 | 110 |
| Σ | 800 | 745 | 2085 |

*RQ2. How do the attitudes expressed by the team covary with software task completion performance?*

Statistical modelling was used to build a model to explain how teams' expressions of attitudes (social, positive, negative, cognitive, work, and achievement), and the other variables considered (to accommodate group structure, situation-related and interaction-related factors), were related to task completion performance. Given that the distributions for the six linguistic dimensions were all skewed (as noted above), we first performed Kendall Taub correlation tests to examine whether the different variables considered in Section 3.2 were correlated with the linguistic dimensions. These results are presented in Table 7. Of the correlations computed, the most notable relationships are: negative expression increased when larger teams were undertaking software tasks, and teams expressed more social and negative expression when they communicated more (and longer) messages. These latter results are all medium-strength, statistically significant positive correlations (refer to Table 7 for details). (Note that Cohen's classification is interpreted as indicating a low correlation when $0 < r \leq 0.29$, medium when $0.30 \leq r \leq 0.49$ and high when $r \geq 0.50$ [105].) In addition, of note is that the dimensions for work and achievement, where there was divergence in the two sets of results for the linguistic analysis and directed content analysis (refer to the previous section), did not have any noteworthy association with our other variables.

Given the skewness of our distributions we next performed a natural log transformation on the variables and executed Pearson product-moment correlation tests in order to create our model. Results from this round of correlation analysis were, unsurprisingly, very similar to those in Table 7.

However, these tests were also used to inform the selection of relevant variables for our regression model, in order to ensure that influential variables were included while at the same time avoiding multicollinearity. The only variable removed was message length, as this variable strongly correlated with number of comments and number of developers ($r = 0.83$). That said, when the message length variable was included there was no change in model performance.

Although a statistically significant model did emerge ($F_{9,30,636} = 220.9$, $p < .01$), the Adjusted R-squared value revealed that our model accounted for just 6% of the variance in task completion performance (Adjusted R-squared = 0.061). The *Beta Coefficients* for the significant variables ($p < .01$) in the order of importance are: work = −0.097, cognitive = 0.096, comment count = 0.095, negative = −0.071, achievement = 0.051, positive = −0.048, social = 0.041, number of roles = 0.040 (refer to Table 8 for further details).

Here we see in Table 8 that when teams expressed more *work*, *negative* and *positive* utterances software tasks were developed slightly faster (variances for the three dimensions being 9.7%, 7.1% and 4.8% respectively). On the other hand, more *cognitive* utterances, *comments*, *achievement* and *social* utterances, and roles, resulted in slightly delayed task completion performance (variances from the five dimensions being 9.6%, 9.5%, 5.1%, 4.1% and 4% respectively). In terms of the control variables, Table 8 shows that of those related to group structure, only number of roles was a significant predictor. For the interaction factors considered, number of comments was the only significant variable.

Overall, these two variables improve model performance by 2% (i.e., the Adjusted R-squared for our base model considering only attitudes was 0.041). These results, along with those above, suggest that, although teams expressed different attitudes when undertaking different forms of software task, their expression of attitudes and the other variables considered were not sufficiently significant predictors of software task completion performance. We discuss these results in relation to theory next.

## 5. DISCUSSION

This work explores the possible relations between task differences and team attitudes, and how the attitudes expressed by software teams covary with software task completion performance. We used linguistic and directed content analysis techniques to examine development artefacts from Agile distributed teams. Although our results allow us to make conjectures only, given that they are drawn from one particular software development context, the evidence provided shows that the teams studied expressed different attitudes when working to complete different forms of software task. On the other hand, teams' expressions of attitudes did not bear a major relationship with their task completion performance. We examine these issues in detail in the following discussions.

*RQ1. What attitudes do team members express when undertaking different forms of software task?*

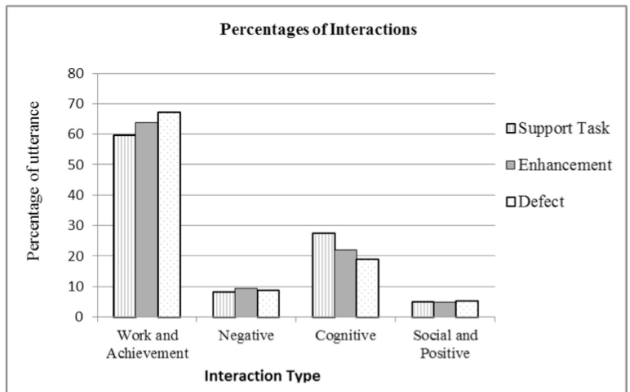

Fig. 3. Percentages of attitudes expressed for sup- port, enhancement and defect tasks.

Our evidence somewhat supports earlier assessments [15,21], that teams express different attitudes when working on different forms of (software) task. We observed that teams expressed elevated levels of both *positive* (using words such as "beautiful", "relax", "perfect" and "proud" ) and *negative* (using words such as "hate", "suck", "dislike" and "stupid" ) attitudes when working to resolve defects, compared to their engagements when undertaking support tasks and enhancements. Previous evidence has found that during software project execution, developers often consider defects to be team obstacles [106], and so, the elevated level of *negative* attitudes that was expressed among teams completing such tasks may be as a consequence. Additionally, given that defects are frequently discovered after features (both new features and enhancements) are coded, and primarily during testing, specific teams working on these defects may already have a sound understanding of the necessary workings to address such issues. Accordingly, teams addressing defects may not necessarily need to express large amounts of *cognitive* (and other) attitudes.

Furthermore, we suspected that more effort would be expended identifying bugs than in applying the fix itself. Thus, there would be less need for a large amount of knowledge exchange, and the expression of emotions more generally, when fixing bugs. In slight contradiction of this assessment, however, in addition to *negative* expressions, we also observed elevated levels of positive attitudes being expressed when teams were resolving defects. High levels of positive expression are indicators of *positive* team climate [107], whereas *negative* emotion is linked to anger and a more cynical team outlook [37]. Both forms of expression are linked to a more emotional state or demeanour. While the relatively high levels of expression of *positive* language by those working on defects is encouraging for team morale and satisfaction [14,25], this could also reflect teams' efforts to offset their frustration and the higher degree of undesirable *negative* attitudes while working on these tasks.

An alternative reasoning could also be that the elevated use of *negative* language when teams were resolving defects was intrinsically part of a defect resolution 'culture' . Previous work has found that software teams working on Mozilla, Eclipse and JBoss used words such as "crash", "critic", "broken", "major", "failure", "error", "trivial", "invalid" and "null" to tag or describe bugs [13]. We therefore manually inspected the negative category in the LIWC dictionary for these words and found that "fail" (not "failure" ) was the closest word in this list that is considered under the *negative* emotion category. Words in the negative emotion category include words such as "bored", "hate", "distress", "suck", "dislike", "angry", "fear", "mess", "stress", "nag", "tense", "problem", "unhappy", and "stupid" . We draw from this that the negative emotion evident here could be more than just reflecting specific terms commonly used around bug fixing, especially given that messages analysed around the different tasks were contributed by the same individuals. In fact, previous evidence shows that developers also expressed negative sentiments around security-related issues [58], suggesting that developers are less pleased working to remedy certain issues. This phenomenon requires further investigation.

Of note is that our small-scale directed content analysis confirms this higher level of negative attitude for teams working around defects. That said, those working to resolve enhancement requests (new soft- ware functionality or the extension of established system features) expressed similar levels of this form of attitude. While similar concerns may be expressed in relation to this finding, we note that only 11.6% of teams' time was dedicated to providing software enhancements, compared to 76.1% for defects.

We articulated in Section 1 that teams addressing defects would likely require higher degrees of familiarity and specific problem-solving knowledge (of the previously

Table 7. Kendall Tau-b Correlation (τ) results for dependent, independent and control variables.

| Other variables | Attitudes | | | | | |
|---|---|---|---|---|---|---|
| | Social | Positive | Negative | Cognitive | Work | Achievement |
| ***Task completion performance*** | | | | | | |
| Time taken | 0.12* | 0.03* | 0.04* | 0.12* | -0.03* | 0 |
| ***Group structure*** | | | | | | |
| No. of developers | 0.28** | 0.26** | 0.29** | 0.17* | 0.04* | 0.07* |
| Number of roles | 0.23** | 0.23** | 0.24** | 0.14** | 0.02** | 0.05* |
| Team lead | 0.15* | 0.09* | 0.14* | 0.08* | 0.05* | 0.06* |
| Admin | 0.01* | 0.05* | 0 | 0 | 0 | 0 |
| Project manager | 0.10* | 0.06* | 0.08* | 0.06* | –0.02* | 0.02* |
| Programmer | 0.10** | 0.15** | 0.15** | 0.07* | 0.01* | 0.02* |
| Multiple | 0.08** | 0.07** | 0.10** | 0.07** | –0.01 | 0 |
| ***Situation-related factors*** | | | | | | |
| Iteration | 0.03* | 0.04* | 0.03* | 0.03** | 0.02** | 0 |
| Priority | 0.05* | 0.02* | 0.06* | 0.05* | 0.01* | -0.01* |
| ***Interaction-related factors*** | | | | | | |
| No. of comments | **0.30**** | 0.25** | **0.33**** | 0.20* | 0.07* | 0.10* |
| Message length | **0.35**** | 0.16** | **0.33**** | 0.28** | 0.04* | 0.07* |

*Note:* **bold** values = medium + ve correlation, ** = p < .01, * = p < .05.

Table 8. Results from regression analysis.

| Variables | | Unstandardized coefficients | | Standardized coefficients | t | Sig. (p) |
|---|---|---|---|---|---|---|
| | | B | Std. Error | Beta | | |
| Attitudes | (Constant) | 0.59 | 0.02 | | 29.52 | 0.00 |
| | Social | 0.09 | 0.02 | 0.041 | 6.10 | 0.00 |
| | Positive | -0.09 | 0.01 | -0.048 | -8.07 | 0.00 |
| | Negative | -0.22 | 0.02 | -0.071 | -12.16 | 0.00 |
| | Cognitive | 0.17 | 0.01 | 0.096 | 13.54 | 0.00 |
| | Work | -0.20 | 0.02 | -0.097 | -12.71 | 0.00 |
| | Achievement | 0.11 | 0.02 | 0.051 | 6.77 | 0.00 |
| Group structure | Number of roles | 0.20 | 0.04 | 0.040 | 5.46 | 0.00 |
| Interaction-related factors | No. of comments | 0.23 | 0.03 | 0.095 | 8.21 | 0.00 |

developed feature), and so, work on such features may require less cooperation, tending towards smaller groups working with increased intellectual processes [25]. Our results are relevant to this assessment, as we indeed observed a reversed pattern for cognitive attitudes, which appeared most pronounced when teams were completing support tasks (design documents and documentation) and enhancements (new functionalities and feature extensions).

We anticipated that coding a new feature or effort spent on feature enhancements would necessitate high amounts of intellectual and cognitive processes [21], and such tasks would present a greater level of difficulty and would require superior levels of idea generation processes [22] than other tasks, an assessment somewhat supported by our evidence. Therefore, confirmation of higher levels of cognitive attitude among teams when they were undertaking enhancements is fitting given the general need for elevated levels of brainstorming when software teams are delineating new or additional requirements. Teams' dialogue around such tasks would comprise words such as "think", "consider", "determined", "idea" "definitely", "always", "extremely" and "certain" ; potential indicators of scaffolding and idea generation among individual members, and all captured under the cognitive cate- gory in the LIWC tool [35,37]. Teams working to resolve support tasks also engaged in higher levels of these processes. Perhaps, competent members that are inclined to share their ideas would make ideal teammates for addressing support tasks and building new features and extending those features that are already developed? Such questions offer fruitful avenues for further research, aimed at assessing the causal linkages between the sentiments expressed by software teams and their actions during development.

In fact, our linguistic analysis shows that teams used the highest levels of *work-* and *achievement-* utterances/concerns when working to address support tasks (e.g., design documents and documentation). Such processes reflected use of words such as "feedback", "goal", "de- legate", "accomplish", "attain" "resolve" and "finalize" . We anticipated that activities related to documentation, design or software support would demand higher levels of manipulative and cooperative requirements [16]. Such a need for higher levels of cooperation when undertaking requirements gathering and software design has indeed been observed by others examining the collaboration patterns of software developers [23]. Additionally, increased persuasion and consensus [24] may benefit those operating on documentation and software support and design tasks. The higher level of *work-* and *achievement-*focused expression observed, along with relatively high levels of social processes that were seen when teams were working to complete such tasks in this work, support our early proposition. However, our directed content analysis outcomes for *work-* and *achievement-*focus have diverged somewhat from those that were returned from our linguistic analysis. This divergence was also observed for teams resolving defects, where our directed content analysis outcomes differed from the linguistic analysis outcomes for *work-* and *achievement-*focus. Given the smaller sample of messages that were analysed using the directed content analysis approach (1261 of the total 117,101), we are not able to draw further definitive inferences from these outcomes.

In an iterative and Agile development context the cycles of design, code and test are repeated frequently, and so naturally, teams' various expressions (and by extension, their motivations) would almost certainly change over time. Software design work is generally completed prior to coding new features or feature enhancements, whereas defects are typically detected during software testing. Perhaps these Jazz teams were eager to start coding software features, and then subsequently eager for their release in order to undertake other work. Role theories indeed show that individuals and teams who are most motivated to complete their tasks are most task-focused [107]. Such individuals may also be most driven and cognitive. That said, our aim in this work was focussed on investigating the relationship between task differences and team attitudes. We believe that our evidence could encourage future work aimed at understanding further why the patterns noted existed, and their potential consequences for software development teams and their realization of project goals. This is particularly relevant for the software development community in light of recent evidence showing that software developers' performance tended to reduce under conditions where they were unsatisfied [71]. Furthermore, in environments where teams work closely together to develop software, such as in Agile contexts, it would seem crucial to understand the reasons for dissatisfaction in order to maintain team morale.

*RQ2. How do the attitudes expressed by the team covary with software task completion performance?*

Our findings did not reveal a strong link between teams' expression of attitudes and their task completion performance. However, we did observe that when teams expressed more work, negative and positive utterances software tasks were resolved slightly faster. In addition, we observed that larger teams working to execute software tasks expressed slightly more negative attitudes. Furthermore, coinciding with this result, we noticed that teams expressed more social and negative attitudes when they communicated more messages. Although we did not initially establish theoretical support for a relationship between larger teams and heightened emotions, it is plausible that larger teams may indeed experience some form of information transmission and propagation delay [108]. This could in turn result in team members being more emotional. A larger team may also promote information diversity, which may necessitate the need for team members to manage higher volumes of information, thus resulting in heightened emotions [102,103].

Another explanation for our result here may be that in larger teams the need to express emotional content is greater to maintain team balance (a position also noted above). For example, to ensure that group harmony is preserved, individuals may put in a little extra effort on positive sentiment in their communication. Social desirability may also be greater in larger groups, as in small groups it is easier to gauge how the members will perceive a message, whereas in large groups it may be "safer" to exaggerate a bit.

In fact, higher prevalence of negative (and positive to a lesser extent) attitudes had a small positive relationship with task completion performance (and as our outcomes for work and achievement linguistic dimensions diverged with those from our directed content analysis, so we restrict our inferences for these dimensions). Thus, in terms of the evidence considered in this study, such attitudes may seem useful for teamwork, although, there is need for further work to confirm this pattern of results. On the other hand, negative language (e.g., hate, suck, dislike, stupid) may be an indicator of frustration, which may some- times lead to conflicts. While some conflicts are in fact useful for maintaining critical evaluation [109], too much of this form of attitude may be disruptive and has the potential to detrimentally affect team performance [110]. For instance, weaker members may become hesitant to solicit help from more capable and aware colleagues if such members' expressed attitudes that are deemed to be negative (or unfriendly). Previous work has indeed established that team members reduce their communication in team environments that are less friendly [45], and unhappiness has been linked to negativity [72]. On the other hand, groups with higher levels of positive-negative attitude ratios have also been shown to operate with a high level of satisfaction [44] and motivation [14]. Positivity was also linked to enhanced software development teams' performance and overall team happiness [73]. Such an arrangement is particularly useful for Agile teams such as those studied in this work, and especially those working in a distributed development context, where there are limited opportunities for group bonding through face to face contact.

In fact, the cohort of teams studied here is no doubt highly skilful (given the global success of the products emanating from the Jazz projects), and so their expressions of negative attitudes did not seem to adversely affect their performance, and particularly given that they completed tasks faster when this form of attitude was higher. That said, it would be undesirable for such attitudes to prevail in software development team environments where overall performance is highly dependent on members' camaraderie and more social and friendly team norms [50]; for instance, during requirements gathering or when members are jointly working on specific system components.

We observed that when teams communicated more cognitive utterances, comments, and social utterances, and there was a wider spread of roles involved, task completion performance was delayed slightly. We are not sure, however, if the outcomes here are influenced by other factors not measured in this work (e.g., task completion performance and the incidence of the aforementioned sentiments in the communication may be affected by task difficulty). While the outcomes for social utterances and the higher spread of roles were negligible, cognitive utterances and comments had a larger effect. Cognitive messages can be influenced by more exchanges or a more pronounced cooperation requirement [16], and thus, may result in more team effort being spent on communication [22]. This said, it would be undesirable to trade-off exchanges of ideas that may lead to innovativeness [75, 76] for marginal levels of task delays, given the benefits that are derived from innovative thinking (e.g., exploring potential new or better ways of doing things).

# 6. THREATS TO VALIDITY

While we have provided a number of insights in this work we acknowledge that there are a number of shortcomings that may potentially affect the validity and generalizability of our study outcomes, and we consider these in turn:

(1) The LIWC language constructs used to measure **attitudes** in this study have been utilized previously to investigate language use and how the expression of sentiments correlated with various forms of psychological processes (refer to Section 2 and Section 3.2 for discussion around language use and its relation to attitudes). In addition, the LIWC tool's dimensions were extensively assessed for validity and re- liability [35,37,88]. However, although the LIWC dictionary was able to capture 66% of the overall words used by Jazz teams, the adequacy of these constructs in the specific context of software development warrants further investigation. To that end, we checked a small sample of the messages to see what might account for the remaining words being ignored by the LIWC tool and found that there were large amounts of cross-referencing to other WIs in the messages, along with large amounts of highly specialized, technology-related language (e.g., J2EE, LDAP, HTTP, Servlet, WIKI, HTML) evident in Jazz members' exchanges. Their non-consideration here is therefore not a problem, as such terms are linguistically neutral with respect to attitudes. In addition, we triangulated our LIWC outcomes using directed content ana- lysis, and classification schemes that were developed for studying interactions [89,90]. In fact, our reliability assessment measure revealed excellent agreement between coders, suggesting that our findings benefitted from accuracy, precision and objectivity [95].

(2) We computed **performance** by calculating the number of day(s) it took for a **task** to be completed. A task was assessed as completed if the status was set to resolved, closed or verified and a corresponding date added. This measurement has been used previously to assess de- livery performance [79,99]. However, we cannot be certain that team members updated each record in an accurate and timely manner. In addition, there is a possibility that inherent differences in task complexity and size may have influenced our performance outcomes. All software tasks are not equal, and especially given the intangible nature of software development [29,30,32]. For instance, fixing a bug may simply require correcting a syntax error in one instance, whereas in another instance re-engineering a class or method may be required. In addition, a software tasks may be deemed completed today, and the software feature released, only for users to find bugs months later and the feature to need maintenance or repair. Such tasks would be incorrectly labelled as completed given the need for new work. That said, others have assessed the Jazz data as generally representative of the project's realities [6], and so it offered us an opportunity to explore a useful area of software engineering human factors.

(3) **Communication** and teams' **interactions** were assessed based on messages sent around explicit software tasks. These messages were extracted from Jazz, and so, may not represent all of the project teams' communications. In fact, while some members were collocated, others were not. Collocated members are likely to engage outside of the Jazz environment, and these engagements are not easily captured for ana- lysis. Offsetting this concern is the fact that, as Jazz was developed as a globally distributed project, teams were required to use messages so that all other contributors (irrespective of their physical location) were aware of product and process decisions regarding each WI [9]. To this end, we anticipated that a significant amount of the teams' communications was captured in our analysis.

(4) A **single organization** employing particular Agile development practices was examined in this study. Work processes and work culture at IBM are likely to be specific to that organization and may not be representative of organization dynamics elsewhere, and particularly for environments that employ conventional waterfall processes. Such environments may employ more rigid project management practices, with much clearer hierarchical structures, development boundaries and other defined roles [17]. In fact, we have not examined actual team meetings and discussions in this work, which affects the richness of the evidence we provided above. That said, Costa, Cataldo, and de Souza [111] confirmed that teams in the Jazz project exhibited similar co- ordination needs to those of four projects operating in two distinct companies. Thus, we believe that our results may be applicable to similar large-scale distributed projects.

# 7. IMPLICATIONS

Although we did not observe a strong link between teams' expressions of attitudes and their task completion performance, we contend that multiple patterns noted in our results have implications for re- searchers studying the attitudes of software teams and for those governing software projects. We consider these issues in the following two subsections.

**7.1. Implications for theory**

While a wealth of research has examined the feasibility of predicting the incidence and resolution of bugs given the way software developers describe such tasks [10–13], less effort has been dedicated to understanding how software teams express attitudes across the full range of software tasks that are commonly performed. Notwithstanding the volume – and therefore the importance – of defects that are incurred and commonly detected in proportion to the other software tasks that are undertaken, considering how teams perform across all software tasks would help us to more comprehensively understand software teams' attitudes. This is particularly necessary for environments that stress the value of team collaboration, such as Agile development. Theories from other disciplines have indeed provided a wealth of evidence in support of the requirement for understanding teams' and individuals' interactions and performance across the full range of tasks that are commonly undertaken, in order to achieve a broader understanding of both task differences and team performance [14,15].

Insights from such broader coverage are useful for advancing theory, in a domain where there is a shortage of theoretical understandings [112]. For instance, previous work has noted that software developers perceive defects to be obstacles [106]. We have observed in this work that the Jazz teams expressed heightened emotion when resolving defects, in a way supporting this prior assessment. However, questions around the specific properties of defects that make teams emotional, and/or whether specific interventions (e.g., employing more rigorous testing procedures or code reviews) may reduce defects, and thus, members' emotional dispositions, remain. In fact, we are not sure if it is the incidence of defects that resulted in the heightened level of emotion, and so we encourage future work to further explore this phenomenon. Such work may build on our findings by modelling negative attitudes against bug fixing and team frustration, on the basis that some software features generate less optimism [58]. Negative emotions were also linked to programming tools and languages [61], thus, more granular theoretical models may be proposed in forming the basis of future work that may lead to solid theories (e.g., fixing defects that involve specific third-party APIs increases negative attitudes and frustration which in turn increases delays). This proposition somewhat supports our own outcomes in this work, where we observed that teams expressed significantly lower levels of cognitive attitudes and work and achievement focus when resolving defects compared to when they were working on the other forms of task. While we anticipated that more effort would be expended identifying bugs than in applying the fixes themselves, and so there would be less need for a large amount of knowledge exchange when fixing bugs, this finding is also somewhat supportive of those uncovered previously about developers' limited motivation when resolving defects, and our theoretical proposition above. Furthermore, we are not entirely certain about what specific elements of design documents, documentation, and coding tasks intensified teams' cognitive focus and their drive to complete such tasks. Further enquiries into these issues would provide thought-provoking insights for the knowledge base on software engineering human factors. There is sufficient evidence in this work to propose that teams express different attitudes when working to address various forms of software tasks, and these attitudes are likely to have a bearing on how they function as a group.

Our findings relating to the way in which larger teams expressed more *negative* attitudes, and how elevated levels of both social and negative attitudes were evident for teams that communicated more messages, are also insightful, and could have implications for future work. While we believe this evidence could be linked to the challenges that arise with information propagation overhead and information diversity [101–103] (or a deliberate drive to comment using such utterances), our findings may also be related to multiple additional factors (e.g., the effects of the specific mix of personality traits [113]). Additionally, while we did not observe a strong link between teams' expressions of attitudes and their task completion performance, this outcome may have been affected by the skill-sets of the team members that we observed (and we were unable to consider this variable in our study due to the unavailability of the relevant data). Similarly, the adequacy of the LIWC constructs may also be questioned. In particular, we would encourage follow-up work to focus on further exploring the way the expression of negative attitudes impacts team synergies and norms during various forms of development activities, taking developers' expertise into consideration (perhaps through the use of inductive analysis techniques). Research may also use other forms of sentiment analysis techniques for triangulation. We hope to peruse such directions, and also consider how understandings from actual team meetings may triangulate our outcomes. Insights from such investigations would likely be useful for further understanding the effects of negative attitudes on teamwork, for different forms of teams. These insights would extend the software engineering knowledge base around teams' behavioural processes, and provide useful support for aiding with software project governance. We examine this latter context next.

### 7.2. Implications for practice

Software development, and especially when conducted using Agile approaches, remains a predominantly human-centric activity, undertaken by individuals and in teams. Thus, efforts aimed at providing insights into the way teams work to deliver software are noteworthy for enlightening those in charge of software project governance. Understandings of team processes are particularly useful for teams that place a high level of emphasis on individuals and interactions over processes and tools (as those studied here, [31]), where issues related to team dynamics may drive or derail team success. In fact, there is value in understanding how software development teams work more generally, and the contexts under which they are most likely to perform, regardless of the software development method(s) that are utilised. The resulting insights would help team leaders to understand the complexities in team attitudes that could better inform team composition strategies [114,115]. Outcomes in this study provide contributions to this cause.

For instance, given the large number of defects addressed relative to the lesser volumes of other tasks that were resolved in the Jazz projects, it is reasonable to infer that reducing the incidence of defects would free up substantial additional time for developing and delivering new features or work items. Thus, team leaders observing such patterns may leverage approaches that seek to reduce the effort expended on defect resolution. Mechanisms for finding duplicate bug reports in repositories [116], or those that eliminate or lessen the need to resolve defects in the first instance, such as test-driven development (TDD) or the practice of continuous integration (CI), could go some way to aiding this cause [117]. The latter two interventions are often used in Agile environments, albeit with varying level of strictness [29]. Software development project managers may use these finding to encourage the adoption of such practices more widely during the execution of their software projects (e.g., for projects taking on a hybrid tone).

In fact, our findings suggest that interventions aimed at reducing the incidence of defects could also indirectly affect software teams' attitudes and team synergies. On the premise that an emotional state may not be beneficial for teamwork, managers and team leaders may look to implement activities that encourage a relaxed team atmosphere (e.g., social activities). That said, other strategies may also be directly implemented to improve teams' climate. For example, the heightened emotion that was expressed when teams were resolving defects suggest that a team strategy aimed at rotating those assigned to bug-fixing could reduce frustration. Another direct strategy that may maintain desirable levels of positive-negative attitude ratios, and so potentially enhance confidence among teams, is to assign bugs to more confident, prudent and constructive teams [70]. We note, however, that Agile methods do not recommend specific teams for development and specific teams for debugging, but rather, all teams are meant to take responsibility for the whole development cycle. This does not necessarily limit the option to identify and encourage project champions for major milestones, including testing and those aspects that may introduce team frustration. For instance, while all teams may participate in requirements capture, design, development and testing, specific highly skilled and prudent members may take oversight of those issues that are recurrent, where less skilled members become stuck and time pressure is beckoning. Such a move is likely to generally increase team satisfaction and improve collaboration processes. Given that software teams spend perhaps the majority of their work time resolving defects, this increased satisfaction could likely translate into more desirable team norms and a friendlier team atmosphere, which may have a positive effect on team morale.

Also of note, however, is that, overall, we did not find that negative attitudes were linked to delays in software task completion. In fact, we found that tasks took less time when the expressions of both negative and positive attitudes were higher among teams. Thus, while negative attitudes may threaten team togetherness, there is also evidence that such attitudes, in conjunction with those that are positive in nature, could be constructive for teamwork. Similarly, there may also be trade-offs between teams engaging in more cognitive processes and higher levels of communication and low levels of task delays. Thus, managers should accept that teams may take longer to deliver on task outcomes when there is more cognitive load and more frequent communications to peruse. The benefit to this off course could be realised at a later stage in terms of the delivery of innovation. These are insights to consider when a project manager is observing group dynamics.

Finally, notwithstanding the single case that is considered in this work, our findings also suggest that maintaining smaller teams and encouraging communication brokers to assist with knowledge dis- semination may go some way towards improving team synergies. Project managers and team leaders are thus encouraged to consider these strategies in their management of software teams.

## Acknowledgments

We thank IBM for granting us access to the Jazz repository, and the coders for their help in the data analysis. Thanks also to the reviewers for their very informative comments aimed at improving the earlier versions of this work.